\documentstyle[aps,twocolumn]{revtex}
\input{epsf}
\renewcommand{\baselinestretch} {1.30}

\def\a{\alpha}
\def\b{\beta}
\def\o{\omega}
\def\s{\sigma}
\def\ur{\uparrow}
\def\dr{\downarrow}
\begin{document}
\title{Comment: on possible time reversal symmetry breaking in normal state of
ruthenate.} 

\author{ A.A.Ovchinnikov$^{1,2}$, ~~M.Ya.Ovchinnikova$^1$}
\address{\it $^1$Joint Institute of Chemical Physics, Moscow.\\
$^2$Max Planck Institute for Physics of Complex Systems, Dresden.\\}


\maketitle
\begin{abstract}

A spiral spin structure with nesting vector is expected to be the ground
normal state of $Sr_2RuO_4$. It might be observed with use of the  ARPES
method with circularly polarised photons designed in \cite{1}.

\end{abstract}


The latest exciting results of A.Kaminski et al. \cite{1} on the polarisational
photoemission study of cuprates allows to propose another  object with time
reversal symmetry breaking 
for similar study. It seems to be interesting to apply the same ARPES method
to study the polarisation phenomena in a single-layered ruthenate $Sr_2RuO_4$.
It attracts an attention as a superconductor  
($T_c\sim 1.5K$)  with possible triplet type of pairing \cite{2,3}. 
The main argument in favour of such type of pairing is the Knight shift
behaviour \cite{4,5}.  Several models for a symmetry of the triplet
superconducting (SC) 
order have been discussed \cite{3,6}. Experimental studies ( NQR
relaxation rate , heat capacity, anisotropy of thermal conductivity
 and so on, see \cite{7} and references therein) point to an existence of the
node lines in the superconducting gap.  It  does not exclude the conventional
d-wave pairing also.  

Unconventional  triplet SC  order breaks a time reversal symmetry. But the
symmetry 
breaking may occur in normal state also. Recently we  suppose \cite{8}
that  the normal state of $Sr_2RuO_4$ has a spiral spin structure. The mean
field (MF) study shows that  the energy of spiral state is lower than that of
the  para-, ferro- and antiferro-magnetic states if the spirality vector $Q$
coincides with a nesting vector $Q=2\pi({1\over 3},{1\over 3})$ of $\a,\b$
bands of the ruthenate \cite{9,10}. This hypothesis is consistent with an
incommensurate peak observed 
in inelastic neutron scattering at $q\sim (0.3,0.3,1)$ in units $({2\pi \over
a},{2\pi \over a},{2\pi \over c})$ \cite{11}. But a spiral structure means that
the normal state 
is a state with a broken time-reversal symmetry. This is a state with the
non-zero spin currents ${\bf J}_{\ur}=-{\bf J}_{\dr}\sim -{\bf Q}$ of opposite
directions for two different spin polarisations perpendicular to the
spin-rotation plane of the spiral state. It means that electrons with up
(down) polarisation occupy preferentially the k-states with ${\bf k}{\bf Q}<0$
or  ${\bf k}{\bf Q}>0$ correspondingly. Such asymmetry leads to the
polarisation 
asymmetry of Fermi surfaces found in MF solutions in \cite{8}. It has been
shown that in spiral state the contributions $A_{\s}(k,\o)$  to the total 
one-particle spectral function $A(k,\o)$ from different spin polarisations
$\s=\ur$ or $\dr$  display at $\o\to 0$  the different sections of the Femri
surfaces (see Fig.2 in \cite{8}). So a strong dependence of the photoemission
signal on a direction of a circularly polarised light should be observed. The
100\% effect for the long ranged spiral order is expected to be greatly
suppressed by any domain structure or any sort of a disorder. In particularly,
it must be destroyed by the thermal cycling procedure, which removes a
set of shadow Fermi surfaces from the ARPES map of $Sr_2RuO_4$ \cite{12}.

Work is supported by Russian Fund of Fundamental Research (Projects
No. 00-03-32981 and No. 00-15-97334). Authors thanks V.Ya Krivnov for useful
discussions and P.Fulde for possibility to work in
Max Planck Institute for Physics of Complex Systems, Dresden.

\renewcommand{\baselinestretch} {1.00}

\end{document}